# Magnetotransport studies of FeSe under hydrostatic pressure

Brajesh Tiwari, Rajveer Jha and V.P.S. Awana*

CSIR-National Physical Laboratory, Dr. K.S. Krishnan Marg, New Delhi-110012, India

**Abstract:**

The discoveries of iron-based superconductors with relatively high transition temperature are under intense experimental and theoretical investigation. Here we present magnetotransport measurements on FeSe superconductor under hydrostatic pressure. We show that in Fe-deficient tetragonal FeSe binary compound, the onset of superconducting transition is almost doubled under 1.98GPa pressure and the estimated upper critical field of 26.7Tesla is increased to 47.5Tesla.

*Corresponding Author
Dr. V. P. S. Awana, Principal Scientist
E-mail: awana@mail.npindia.org
Ph. +91-11-45609357, Fax-+91-11-45609310
Homepage www.fteewebs.com/vpsawana/

## Introduction

Superconductivity in iron-based compounds is fast developing research subject in condensed matter physics and materials science community [1-4]. Since most of the Fe-based compounds show magnetic ordering at low temperatures, the unexpected discovery of 26K superconducting transition in $LaFeAsO_{1-x}F_x$ by Kamihara *et.al.* [1] in 2008 rejuvenate the superconducting research community. Fe-based superconductors share common features with their relatives' high-Tc Cu-based (curates) superconductors for example, both are d-electron materials having layered structure, their parent compounds are antiferromagnetic and superconductivity arises from electron or/and hole doping or by application of pressure. The detailed review on this topic can be seen in Ref 2 and Ref 3 [2, 3]. A simple binary compound FeSe in tetragonal phase has emerged as a superconductor with $T_c$ of ~ 8-13K [4-7]. Essentially FeSe has identical packing to that of family of layered FeAs-based high-$T_c$ superconductors however lacking intercalated charge reservoir layers [8]. Iron arsenide and iron selenide share common structure which contains $Fe_2X_2$ (X= As and Se) layers of edge-shared $FeX_4$ tetrahedra suggesting same mechanism responsible for superconductivity in FeSe. In contrast to FeSe, chemical substitutions require for LiFeAs [9] $BaFe_2As_2$ [10] LnFeAsO [1, 2] to drive them to superconductivity regime from itinerant antiferromagnetic state. Superconductivity in FeSe is very sensitive to stoichiometry, defects and disorder in the system [11]. In iron selenide, using various cationic spacer layers consisting of metal ions (for example $Li^+$, $Na^+$, $K^+$, $Ba^{2+}$, $Ca^{2+}$, $Eu^{2+}$ etc), the superconductivity could be increased to $T_c$ of upto 45K [12, 13]. Also using different molecular spacer layers, the superconducting transition is achieved as high as 56K in FeSe based superconductors [14, 15]. The very first report on superconductivity of FeSe under pressure of 1.48GPa was reported by Mizuguchi et.al, with increase of $T_c$ from 13.5K to 27K [16]. The detailed pressure evolution of low temperature structural studies revealed an intimate link between superconducting properties and the Se-FeSe interlayer separation [17]. The nuclear magnetic resonance (NMR) measurements by Imai *et. al.* suggested link between spin fluctuations and superconductivity in FeSe, however the Mossbauer studies confirmed that the same cannot be attributed to suppression of spin-density wave, unlike arsenide superconductors [7, 18]. In the present work, we revisited the superconductivity in FeSe under hydrostatic pressure via magnetotransport measurements. These measurements were performed on Fe-deficient tetragonal FeSe sample with minor hexagonal (NiAs-type) impurity.

**Experimental details**

Polycrystalline bulk samples were prepared from high purity (>99 %) powder of iron and selenium mixed and ground in an agate mortar and pestle with nominal stoichiometry of FeSe. The grinding was carried out in argon filed glove box. The grinded powder was cold pressed into rectangular pellet with 100kg/cm$^2$ uniaxial stress. The pellets were sealed in an evacuated quartz tube (in $10^{-4}$ bar vacuum) and temperature was ramped in a box Furness to 750$^o$C i.e., above boiling point of Se (685$^o$C) at the rate of 2$^o$C/min. The sample was kept at this temperature for 12h before cooling to room temperature by rate of 2$^o$ C/min. The sample was re-grinded and pelletized before final sintering at 750$^o$C for 12h. Powder x-ray diffraction (XRD) data was recorded at room temperature in 2θ range of 5$^o$ to 60$^o$ in step of 0.02 using Rigaku made diffractometer with Cu K$_α$ ($λ$ = 1.54Å). Multiphase Rietveld refinement was carried out using Full-Prof software. Physical Property Measurements System (*PPMS*-14T, *Quantum Design*) is used to perform magnetotransport measurements under hydrostatic pressure, which is applied using HPC-33 Piston cell in addition to DC resistivity Option. Hydrostatic pressures were generated by a BeCu/NiCrAl clamped piston-cylinder cell under a fluid medium of Fluorinert in a 3 Teflon cell. Annealed Pt wires attached to gold-sputtered contact surfaces on the sample by silver epoxy in a standard four-wire configuration. The applied pressure was calibrated from the superconducting transition temperature of Pb.

**Results and Discussion**

X-ray powder diffraction pattern is Rietveld refined together for tetragonal P4/nmm and hexagonal P6$_3$/mmc structures of FeSe using Full-Prof and is shown in figure 1(a). The global fitness of XRD pattern is $χ^2$ = 3.21. Nonsymmorphic tetragonal structure of FeSe with space group P4/nmm (#129) is best fit ted to the lattice parameters a=b=3.774(2)Å and c=5.520(6)Å along with hexagonal (NiAs type) P6$_3$/mmc (#196) as a minority phase (with lattice parameters a=b=3.626(5)Å and c= 5.890(1)Å), which is close to the previous reported values [6, 8, 11, 18]. Both tetragonal and hexagonal crystal structures are represented in figure 1(b). In tetragonal phase, Fe and Se occupy the Wyckoff positions; 2a (3/4, 1/4, 0) with site symmetry -4m2 and 2c (1/4, 1/4, 0.275) with site symmetry 4mm, respectively. However in hexagonal phase Fe occupies Wyckoff position; 2a (0,0,0) with site symmetry -3m and Se; 2c(1/3, 2/3, 1/4) with site symmetry -6m2. Tetragonal phase of FeSe is formed by layers of FeSe whereas hexagonal phase does not have this type of structure. The best fit of XRD pattern of FeSe gives Bragg R-factor of 3.45 and 3.28 for tetragonal and 20% of secondary hexagonal phases, respectively and deduced stoichiometry Fe$_{0.954\pm0.005}$Se in its tetragonal phase. When we forced the software to strict stoichiometry of FeSe, the obtained Bragg-R factor and reduced $\chi^2$ are higher compared to the composition Fe$_{0.954\pm0.005}$Se and hence confirms the Fe-deficient FeSe in its tetragonal phase. This Fe-deficient tetragonal phase of FeSe shows superconducting transition which is consistent to the recently reported compositions by Chen *et.al.* [19] but in contrary to some earlier reports [5-8].

Resistivity (ρ) measured under different external hydrostatic pressure upto 1.98GPa is displayed in figure 2 for the studied FeSe. Normal state resistivity follows the metallic behavior for FeSe, as can be seen in figure 2(a). Interestingly, an anomalous deviation under increasing pressure can be observed in resistive ty at above ~100K. Based on NMR investigations Imai *et.al.* suggested that at below 100 K temperature, antiferromagnetic spin fluctuations are strongly enhanced towards superconducting transition [7]. Magnified part of ρ-T curve can be viewed as inset of figure 2(a) from which onset of T$_c$ is determined by extrapolating the curve to straight lines (solid lines). The obtained T$_c$ onset and offset under different pressure is presented in figure 2(b). Superconductivity is observed in FeSe below found to increase by application of pressure in contrast to conventional superconductors and reaches to 20.2K under 1.98GPa, accompanying broadening of transition. However the offset transition temperature which is 6.2K in ambient pressure is almost doubled under hydrostatic pressure of 1.98 GPa. Due to broadening of transition upon application of higher pressures the onset and offset temperature deviates. This rather broad transition width indicates the inhomogeneous nature of the sample. The change of T$_c$ under pressure may be related to softness of Fe$_2$Se$_2$ planes as shown in figure 1(b), which have significant influence on electronic properties. It was noticed based on Mossbauer study by Medvedev et.al that local environment around Fe is not much affected by pressure [18].

To further study the electronic transport of FeSe under hydrostatic pressures of 0 GPa and 1.98 GPa, resistivity measurements as a function of temperature were carried out under different applied magnetic fields parallel to the current flow direction. Under ambient pressure (0GPa) magnetotransport measurements is displayed in figure 3(a) where resistive transition shifts towards lower temperature upon increasing magnetic field with broadening of transition width. Under 1.98GPa hydrostatic pressure the resistive transition as displayed in figure 3(b) does not show an increase in transition width with applied magnetic field in contrast to 0GPa case. This suggests a clear enhancement of flux pinning in FeSe sample under applied pressure.

Shown in figure 4 is the temperature dependence of fields $H_{c2}^{90\%}$ and $H_{c2}^{10\%}$ evaluated at 90% and 10% of normal state resistivity, $\rho_N$. The temperature dependence of upper critical fields are very close to linear allowing us to use the Werthamer Helfand and Hohenberg (WHH) model for the determination of uppercritical field $H_{c2}(0)$ by $H_{c2}(0) = -0.69 T_c \left(\frac{dH_{c2}}{dT}\right)_{T=T_c}$ relation within weak coupling BCS superconductors [20]. The rate of decrease of upper critical field, $dH_{c2}/dT$, using the 90%$\rho_N$ and 10%$\rho_N$ criteria are -3.69Tesla/K and -3.00Tesla/K which decrease to -3.40 Tesla/K and -2.91 Tesla/K respectively under hydrostatic pressure of 1.98GPa. We estimated the upper critical field using WHH formula which surprisingly increases from 26.7Tesla to 47.5Tesla under 1.98GPa pressure. The upper critical field value under ambient pressure is close to earlier reported values [16-19]. In a very recent work by Miyoshi *et.al.*, on single crystal of FeSe specimen, the increase in Tc upto 34 K under hydrostatic pressure of 7GPa is observed, however the does not comment on upper critical field and pinning [21]. The superconducting coherence length $\xi(0)$ at absolute zero can be estimated from $H_{c2}(0) = \frac{\Phi_0}{2\pi \xi^2(0)}$ based on Gingberg-Landau theory where flux quantization $\Phi_0 = 2.07 \times 10^{-15}$ Web. The estimated coherence length decreases under 1.98 GPa pressure from 3.51nm to 2.63nm indicating enhanced robustness. The decreased coherence length and enhanced pinning by hydrostatic pressure in conjugation with increased Tc may lead a way to design superconductors for potential applications.

In summary we have studied superconducting properties of synthesized FeSe via the magnetotransport measurements under different hydrostatic pressures. XRD analysis indicates that the superconducting tetragonal FeSe have Fe-vacancy with a minor hexagonal phase. Under hydrostatic pressure of just 1.98GPa, the superconducting onset temperature increase from 10.4 K to 20.2K with broadening in transition. From the magnetotransport measurements the estimated upper critical field is 26.7Tesla which increased to 47.5Tesla under increased pressure which in turn results in decreased coherence length in conjugation to enhanced flux pinning.

**Acknowledgement:**

Authors would like to thank their Director NPL India for his interest in the present work.

**References:**

[1]. Y. Kamihara, T. Watanabe, M. Hirano, and H. Hosono, *Iron-based layered superconductor La[$O_{1-x}F_x$]FeAs (x = 0.05–0.12) with $T_c = 26K$*, J. Am. Chem. Soc. **130**, 3296 (2008).

[2]. G. R. Stewart, *Superconductivity in iron compounds.* Rev. Mod. Phys. **83**, 1589 (2011).

[3]. E. Dagotto, *Colloquium: The unexpected properties of alkali metal iron selenide superconductors.* Rev. Mod. Phys. **85**, 849 (2013).

[4]. F. C. Hsu, J. Y. Luo, K. W. Yeh, T. K. Chen, T. W. Huang, P. M. Wu, Y.C. Lee, Y. L. Huang, Y. Y. Chu, D. C. Yan and M. K. Wu, *Superconductivity in the PbO-type structure α – FeSe.* Proc. Natl. Acad. Sci. **105**, 14262 (2008).

[5]. E. Pomjakushina, K. Conder, V Pomjakushin, M. Bendele, and R. Khasanov, *Synthesis, crystal structure, and chemical stability of the superconductor $FeSe_{1-x}$*. Phys. Rev. B **80**, 024517 (2009).

[6]. S. B. Zhang, X. D. Zhu, H. C. Lei, G. Li, B. S. Wang, L. J. Li, X. B. Zhu, Z. R. Yang, W. H. Song, J. M. Dai and Y. P. Sun, *Superconductivity of $FeSe_{0.89}$ crystal with hexagonal and tetragonal structures.* Supercond. Sci. Technol. **22** 075016 (2009).

[7].T. Imai, K. Ahilan, F. L. Ning, T. M. McQueen, and R. J. Cava, *Why does undoped FeSe become a high-Tc superconductor under Pressure?* Phys. Rev. Lett. **102**, 177005 (2009).

[8]. S. Margadonna, Y. Takabayashi, M. T. McDonald, K. Kasperkiewicz, Y. Mizuguchi, Y. Takano, A. N. Fitch, E. Suarde and K. Prassides, *Crystal structure of the new $FeSe_{1-x}$ superconductor*. Chem. Commun., issue **43**, 5607 (2008).

[9]. M. J. Pitcher, T. Lancaster, J. D. Wright, I. Franke, A. J. Steele, P. J. Baker, F. L. Pratt, W. Trevelyan-Thomas, D. R. Parker, S. J. Blundell and S. J. Clarke, *Compositional control of the superconducting properties of LiFeAs*, J. Am. Chem. Soc. **132**, 10467 (2010)

[10]. D. Mandrus, A. S. Sefat, M. A. McGuire, and B. C. Sales, *Materials chemistry of $BaFe_2As_2$: a model platform for unconventional superconductivity*. Chem. Mater. **22**, 715 (2010).

[11]. T. M. McQueen, Q. Huang, V. Ksenofontov, C. Felser, Q. Xu, H. Zandbergen, Y. S. Hor, J. Allred, A. J. Williams, D. Qu, J. Checkelsky, N. P. Ong, and R. J. Cava**,** *Extreme sensitivity of superconductivity to stoichiometry in $Fe_{1+\delta}Se$.* Phys. Rev. B 79, 014522 (2009).

[12]. T. P. Ying, X. L. Chen, G. Wang, S. F. Jin, T. T. Zhou, X. F. Lai, H. Zhang and W. Y. Wang, *Observation of superconductivity at 30 K~46 K in $A_xFe_2Se_2$ (A = Li, Na, Ba, Sr, Ca, Yb, and Eu).* Sci. Rep. **2**, 426 (2012).

[13]. Jiangang Guo, Shifeng Jin, Gang Wang, Shunchong Wang, Kaixing Zhu, Tingting Zhou, Meng He, and Xiaolong Chen, *Superconductivity in the iron selenide $K_xFe_2Se_2$ (0≤x≤1.0)* Phys. Rev. B **82,** 180520(R) (2010).

[14]. A. K. Miziopa, E. V. Pomjakushina, V. Y. Pomjakushin, F. von Rohr, A. Schilling and K. Conder, *Synthesis of new alkali metal-organic solvent intercalated iron selenide superconductor with $T_c$ ~ 45 K*. J. Phys: Condens. Matter **24**, 382202 (2012).

[15]. M. B. Lucas, D. G. Free, S. J. Sedlmaier, J. D. Wright, S. J. Cassidy, Y. Hara, A. J. Corkett, T. Lancaster, P. J. Baker, S. J. Blundell and S. J. Clarke, *Enhancement of the superconducting transition temperature of FeSe by intercalation of a molecular spacer layer.* Nature Mater. **12**,15 (2013).

[16]. Y. Mizuguchi, F. Tomioka, S. Tsuda, T. Yamaguchi and Y. Takano, *Superconductivity at 27 K in tetragonal FeSe under high pressure*. Appl. Phys. Lett. **93**, 152505 (2008).

[17].S. Margadonna, Y. Takabayashi, Y. Ohishi, Y. Mizuguchi, Y. Takano, T. Kagayama, T. Nakagawa, M. Takata, and K. Prassides, *Pressure evolution of the low-temperature crystal structure and bonding of the superconductor FeSe ($T_c$ = 37 K).* Phys. Rev. B **80**, 064506 (2009).

[18]. S. Medvedev, T. M. McQueen, I. A. Troyan, T. Palasyuk, M. I. Eremets, R. J. Cava, S. Naghavi, F. Casper, V. Ksenofontov, G. Wortmann and C. Felser, *Electronic and magnetic phase diagram of β-$Fe_{1.01}$Se with superconductivity at 36.7 K under pressure.* Nature Mater. **8**, 630 (2009).

[19]. T. K. Chen, C. C. Chang, H. H. Chang, A. H. Fang, C. H. Wang, W. H. Chao, C. M. Tseng, Y. C. Lee, Y. R. Wu, M. H. Wen, H. Y. Tang, F. R. Chen, M. J. Wang, M. K. Wu and D. V. Dyck, *Fe-vacancy order and superconductivity in tetragonal β-$Fe_{1-x}$Se.* Proc. Natl. Acad. Sci. **111**, 63 (2014).

[20] N. R. Werthamer, E. Helfand, and P. C. Hohenberg, *Temperature and purity dependence of superconducting critical field, $H_{c2}$ III. Electron-spin and spin-orbit effects,* Phys. Rev. **147**, 295 (1966).

[21] K. Miyoshi, K. Morishita, E. Mutou, M. Kondo, O. Seida, K. Fujiwara, J. Takeuchi, and S. Nishigori, *'Enhanced Superconductivity on the Tetragonal Lattice in FeSe under Hydrostatic Pressure'* J. Phys. Soc. Jpn. **83**, 013702 (2014).

**Figures captions:**

Figure 1. (a) Rietveld refinement for powder x-ray diffraction pattern of FeSe in both tetragonal P4/nmm (green bars) and hexagonal P63/mmc (red bars) phases. (b) Crystal structures for both tetragonal and hexagonal phases.

Figure 2. (a) Resistivity of FeSe under different hydrostatic pressures as function of temperature with a clear increase in superconducting transition is shown for clarity as inset. (b) Variation of onset and offset transition temperature with applied hydrostatic pressure.

Figure 3. Resistivity (ρ) verses temperature (T) at different magnetic fields under (a) atmospheric pressure, P= 0.0GPa and (b) P= 1.98GPa.

Figure 4. Variation of upper critical field, $H_{c2}(T)$, with temperature as evaluated from 90% and 10 % of normal state resistivity $T_c$ (50% $\rho_n$) under atmospheric and 1.98 GPa hydrostatic pressure. Solid lines are the guide to eyes.

*Fig. 1*

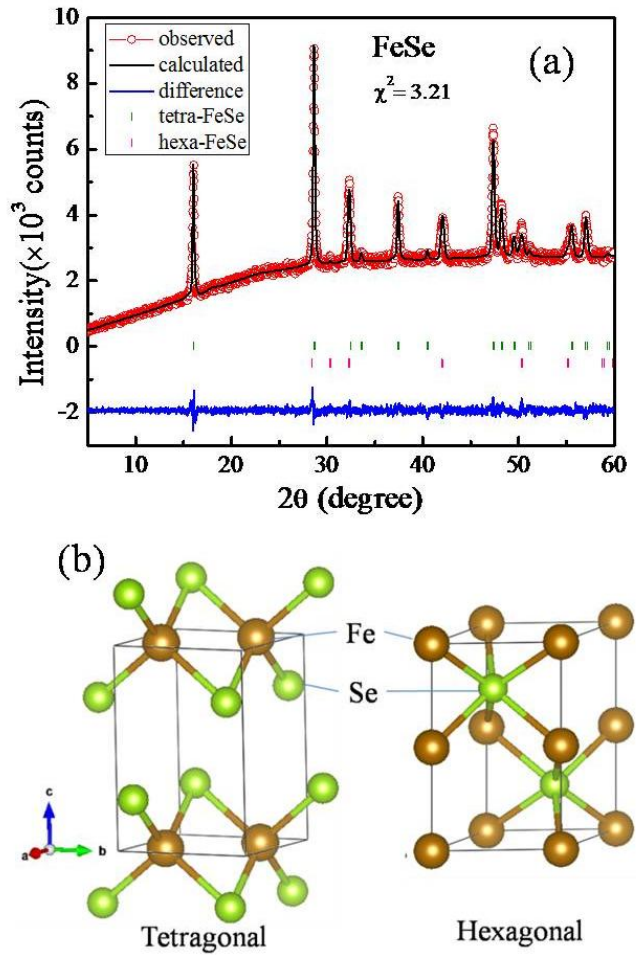

*Fig. 2*            *Fig. 3*

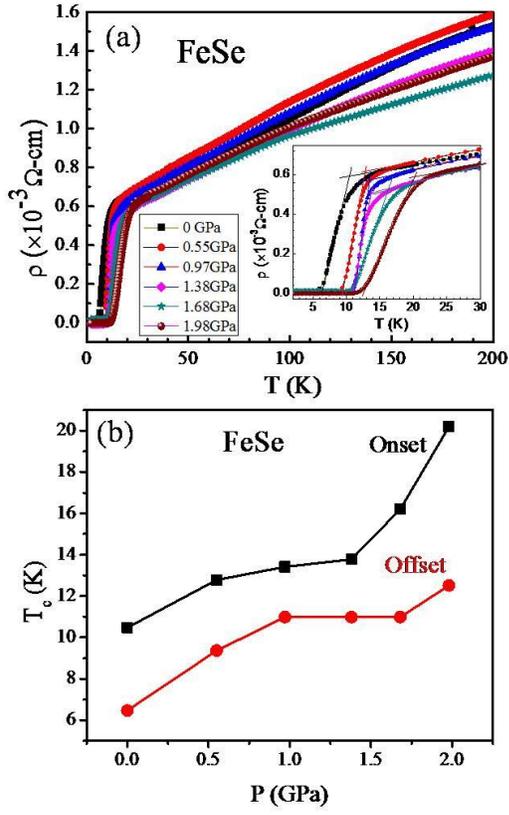
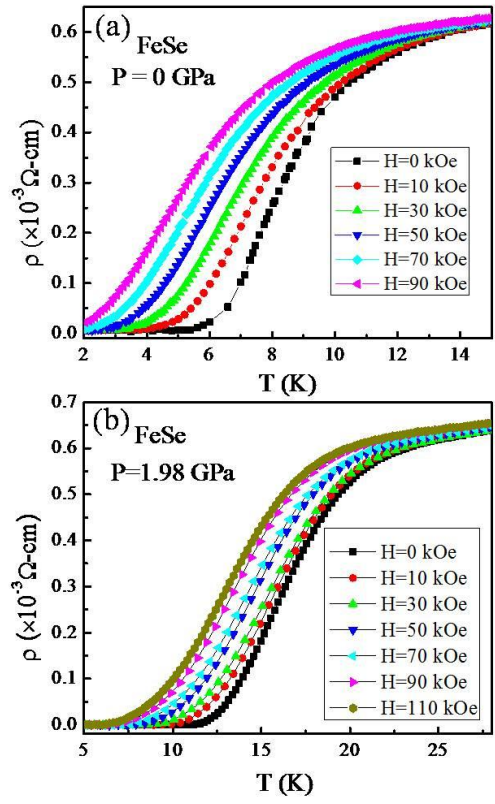

*Fig. 4*

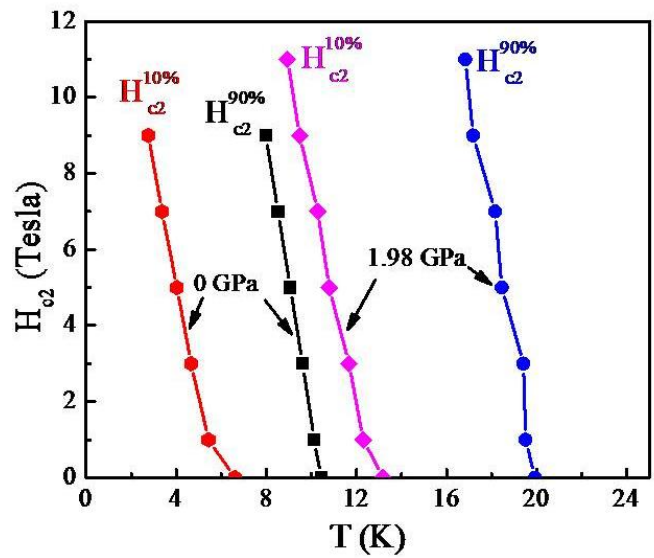